\documentclass[10pt]{iopart}
\usepackage{iopams}  
\usepackage{graphicx}
\usepackage{epstopdf}
\usepackage[colorlinks,citecolor=blue]{hyperref}
\usepackage{cite}
\begin{document}

\title[ECE below the fundamental resonance in ITER]{Modelling the electron cyclotron emission below the fundamental resonance in ITER}

\author{J. Rasmussen$^1$, M. Stejner$^1$, L. Figini$^2$, T. Jensen$^1$,  E.~B. Klinkby$^3$, S.~B. Korsholm$^1$, A.~W. Larsen$^1$, F. Leipold$^1$, D. Micheletti$^2$, S.~K. Nielsen$^1$ and M.~Salewski$^1$}

\address{$^1$ Technical University of Denmark,  Department of Physics, Kgs.~Lyngby, Denmark}
\address{$^2$ Institute of Plasma Physics "Piero Caldirola", National Research Council of Italy, Milan, Italy}
\address{$^3$ Technical University of Denmark, Center for Nuclear Technologies, Roskilde, Denmark}

\ead{jeras@fysik.dtu.dk}
\vspace{10pt}
%\begin{indented}
%\item[]13 June 2019
%\end{indented}

\begin{abstract}
The electron cyclotron emission (ECE) in fusion devices is  non-trivial to model in detail at frequencies well below the fundamental resonance where the plasma is optically thin. However, doing so is important for evaluating the background for microwave diagnostics operating in this frequency range.
Here we present a general framework for estimating the ECE levels of fusion plasmas at such frequencies using ensemble-averaging of rays traced through many randomized wall reflections. This enables us to account for the overall vacuum vessel geometry, self-consistently include cross-polarization, and quantify the statistical uncertainty on the resulting ECE spectra. 
Applying this to ITER conditions, we find simulated ECE levels that increase strongly with frequency and plasma temperature in the considered range of 55--75~GHz. 
At frequencies smaller than 70~GHz, we predict an X-mode ECE level below 100~eV in the ITER baseline plasma scenario, but with corresponding intensities reaching keV levels in the hotter hybrid plasma scenario. Benchmarking against the SPECE raytracing code reveals good agreement under relevant conditions, 
and the  predicted strength of  X-mode to O-mode conversion induced by wall reflections is consistent with estimates from existing fusion devices. 
We discuss possible implications of our findings for ITER microwave diagnostics such as ECE, reflectometry, and collective Thomson scattering.
\end{abstract}

\ioptwocol

\section{Introduction}

As electrons in a magnetized plasma gyrate around the magnetic field lines, they emit and absorb radiation at the local fundamental electron cyclotron (EC) frequency $\omega_c = q_e B/(\gamma m_{e,0})$ and its harmonics at $n \omega_c$, $n>1$. Here $m_{e,0}$ is the electron rest mass and $\gamma = (1-v_e^2/c^2)^{-1/2}$ the relativistic Lorentz factor. For the nominal on-axis toroidal field of $B_t=5.3$~T in ITER, the cold ($\gamma=1$) fundamental EC resonance corresponds to $\omega_c/(2\pi) \approx 148$~GHz.
A number of ITER microwave diagnostics will operate below this frequency, including plasma position reflectometers in the range 15–-75~GHz \cite{rica15}, density reflectometers operating down to similar frequencies \cite{vaya06}, collective Thomson scattering (CTS) for fast-ion measurements at 55--65 GHz \cite{kors16,kors19}, and the electron cyclotron emission (ECE) diagnostic itself down to 70~GHz \cite{tayl15}. In addition, 5~GHz antennas have been considered for external  lower hybrid current drive (LHCD) during long-pulse ITER operation \cite{hoan09,belo15}.

In general, due to relativistic downshift and broadening of the resonances in sufficiently hot ($T_e \gtrsim 10$~keV) plasmas, the ECE contribution to total radiation losses may be significant for optically thin higher harmonics in ITER \cite{born83}. Furthermore, the relativistic downshift of cutoffs at high $T_e$ may render the reconstruction of density profiles  from reflectometry sensitive to accurate knowledge of the temperature profile as determined from, e.g., ECE measurements. The ITER CTS diagnostic could be particularly sensitive to the relativistically downshifted first harmonic emission present in hot plasmas \cite{brus94,born96}, potentially affecting the ability of the diagnostic to measure the distribution function of fusion-born alpha particles. Finally, certain ECE measurements at ITER-like collisionalities in existing tokamaks are themselves hampered by noise, and even by interference from LHCD (e.g.\ \cite{wang17}). 
Application of the above measurement and heating techniques may therefore be impacted by, or even impact on, the ECE levels at $\omega<\omega_c$ in ITER. It thus seems timely to anticipate these ECE levels and their dependence on plasma parameters, in order to characterize ITER plasmas in the optically thin regime and the background radiation expected for diagnostics  at $\omega<\omega_c$. This would aid in preparing both for ITER operation and for fusion devices beyond ITER, in which microwave diagnostics may retain an important role (e.g. \cite{biel15}).

A number of different raytracing codes have been developed for simulations of the ECE in fusion devices, e.g.\ \cite{sill87,chat92,trib96,fari08}. 
Advantages of these codes over simpler analytical and numerical ECE estimates include the possibility to account for effects of refraction, potentially arbitrary viewing geometries and antenna patterns, and the use of realistic electron kinetic profiles in a 3D description of the plasma magnetic equilibrium. Such ECE codes are commonly applied for the study of EC emission and absorption around optically thick harmonics of the fundamental EC resonance (e.g.\ \cite{prat08}) such as the fundamental O-mode or second harmonic X-mode resonances.
It is, however, a challenge to accurately model the ECE emission at frequencies well below the fundamental EC resonance, where the plasma is optically thin, and wall reflections, cross-polarization, and the detailed surface properties of the plasma-facing components (PFCs) may consequently affect the resulting ECE spectrum. 

Given this complexity, it is perhaps not surprising that few predictions for the optically thin ECE spectrum at $\omega < \omega_c$ in ITER seem to be available in the literature. The work of \cite{born96} illustrated how plasmas containing a relativistic electron population can exhibit significant ECE emission associated with relativistically downshifted first harmonic emission at frequencies below the relevant upper cutoff at the outboard plasma edge. However, this was based on models of hot or non-thermal TFTR plasmas in a plane-slab plasma geometry without refractive effects or mode conversion, so accurate extrapolation to realistic ITER conditions is not straightforward. For ITER plasmas, results from the SPECE code \cite{fari08}, limited to frequencies above 100~GHz, have suggested that the above ECE feature may reach peak radiation temperatures comparable to the core electron temperature. However, these results assume a parallel-wall geometry and a pre-specified strength of mode conversion and do not include predictions at lower frequencies. The modelling in \cite{orsi97} (see also \cite{fido96}) does extend down to 60~GHz and suggests significant X-mode ECE levels in ITER above $\sim 70$~GHz, but this is again based on a simplified wall geometry and with no accounting for mode conversion.

To improve on this, we take a statistical approach to this problem by considering an ensemble average of rays traced through an optically thin plasma while subject to many randomized wall reflections in a toroidal geometry. Our technique is distinct from analytical approaches and at least some ECE codes by virtue of its ability to (1) account for the large-scale geometry of the vacuum vessel (allowing generalization to more complex geometries, if relevant), (2) self-consistently compute the level and impact of polarization mode conversion, and (3) quantify and suppress the statistical errors on the resulting ECE spectra using Monte Carlo--based averaging.  Our framework can potentially be applied to any fusion device, but as a specific application, we consider the ECE signal predicted for X-mode waves in ITER  in the frequency range of 55--75~GHz. 

In Section~\ref{sec,ray}, we outline our approach to raytracing of an optically thin plasma, including our technique of ensemble averaging and treatment of cross-polarization and wall reflectivity. Section~\ref{sec,results} presents the simulated ECE spectra and their dependence on model assumptions, and compares our results to other relevant ECE estimates. The
limitations of our approach and possible diagnostic implications of the results are discussed in Section~\ref{sec,discuss}, while Section~\ref{sec,conclude} closes with a summary and outlook.

\section{Raytracing in an optically thin plasma with wall reflections}\label{sec,ray}

In general, a ray travelling through a plasma along a path $s$ from position $s_1$ to $s_2$ will experience a change in intensity or, equivalently, radiation temperature $T_{rad}$. In the Rayleigh--Jeans regime ($\hbar\omega \ll kT_e$) relevant for microwaves, this change is described by \cite{beke66,hart13}
\begin{equation}
T_{rad}(s_2,\omega) =  T_{rad}(s_1,\omega)e^{-\tau_\omega(s_1,s_2)}    
\label{eq,trad}  
\end{equation}
\begin{eqnarray*}
\hspace{2.5cm}  + \int_{s_1}^{s_2} T_e(s) \alpha_\omega(s) e^{-\tau_\omega(s)} \mbox{d}s,
\end{eqnarray*}
%\begin{equation}
%T_{rad}(s_2,\omega) =  T_{rad}(s_1,\omega)e^{-\tau_\omega(s_1,s_2)}    + \int_{s_1}^{s_2} T_e(s) \alpha_\omega(s) e^{-\tau_\omega(s)} %\mbox{d}s,
%\label{eq,trad}  
%\end{equation}
assuming local thermodynamic equilibrium at the electron temperature $T_e$,
where $\tau_\omega(s_1,s_2) = \int_{s_1}^{s_2} \alpha_\omega(s)  \mbox{d}s$ is the optical depth along the ray path, and $\alpha_\omega$ is the frequency-dependent absorption coefficient which depends on the local plasma dielectric tensor. 
Integrating equation~(\ref{eq,trad}) along the ray path, one can thus derive the optical depth and ECE radiation temperature along the ray.

To this end, we use our raytracing code {\em Warmray}, developed by H.~Bindslev. The code includes effects of relativistic electrons on the plasma refractive index \cite{bind92,bind93}, and solves equation~(\ref{eq,trad}) along each ray in the WKB approximation using the weakly relativistic dielectric tensor and plasma susceptibilities of \cite{shka86}.
{\em Warmray}  has been used extensively for the design and interpretation of CTS experiments at TEXTOR and ASDEX Upgrade (e.g.\ \cite{niel17}), providing reliable predictions of the orientation, width, and temporal evolution of microwave beams in numerous CTS experiments. 
Being a raytracing code, it can only give an approximate description of the beam shape by tracing multiple peripheral rays. Nevertheless, benchmarking studies \cite{stej17} against the widely used {\em Torbeam} beamtracing code \cite{poli01} show that both the ray direction and the beam shapes predicted by {\em Warmray} are typically consistent with {\em Torbeam} results, one exception being the case of highly focussed beams for which diffraction effects, not considered by {\em Warmray}, become important. {\em Torbeam} itself has been benchmarked against a range of other ray-- and beamtracing codes \cite{prat08}.

Here we use {\em Warmray} to trace a single ray through many wall reflections and compute the resulting ECE spectrum. This is
in contrast to the "standard" use of the code for CTS experiments, where both a central and four peripheral rays are simulated as a means of approximating the wave distribution in the beam plane. We will briefly discuss the possible impact of this simplification in Section~\ref{sec,discuss}.

For the plasma conditions, we consider two full-field ($B_t = 5.3$~T) D-T operating scenarios: The standard ELMy H-mode ITER baseline plasma scenario ($I_p=15$~MA $Q=10$)
and the longer-pulse hybrid ($I_p = 12.5$~MA, $Q \geq 5$) scenario with higher bootstrap fraction. These are expected to comprise $\gtrsim 90$\% of all D-T pulses \cite{hend15}. Magnetic equilibrium data and electron temperature and density profiles for the H-mode 
flattop phase of these scenarios were extracted from the scenario database incorporated in the ITER Integrated Modelling \& Analysis Suite (IMAS; \cite{imbe15,pinc18}), using the scenario identifiers ({\tt shot,run}) = ({\tt 131007,0}) and ({\tt 130502,1}) for the baseline and hybrid scenarios, respectively. The resulting electron density and temperature profiles are shown in Figure~\ref{fig,cutoffs}, along with the location of the various resonances and cutoffs discussed here.

\begin{figure}
\begin{center}
\includegraphics[width=8.4cm]{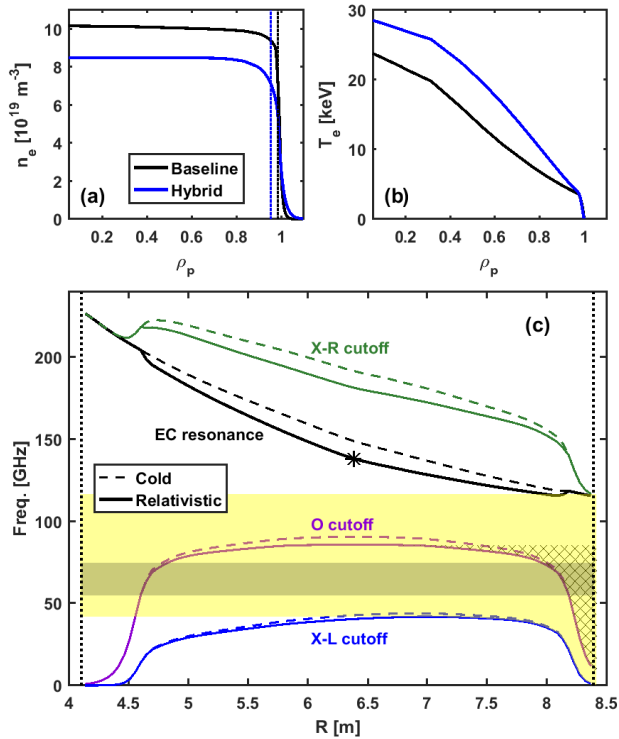}
\caption{Profiles of (a) electron density and (b) electron temperature in the adopted ITER baseline and hybrid plasma scenarios as functions of $\rho_p$ (square root of normalized poloidal flux). Dotted lines in (a) indicate the location of the relativistic 75~GHz cutoff for O-mode waves, computed from \cite{bind93}. Panel (c) shows the location of relevant 
resonances and cutoffs as a function of major radius $R$ in the baseline plasma scenario for wave propagation perpendicular to the magnetic field (results for the hybrid scenario are qualitatively similar). Dashed curves illustrate the cold resonances/cutoffs and solid curves the corresponding relativistic ones based on the characteristic electron thermal energy, with an asterisk marking the location of the magnetic axis. Dotted lines outline the vessel wall in the midplane. The yellow area shows the region from which X-mode radiation below the fundamental resonance can be accessed from the low-field side (LFS), and the hatched area represents the region in which O-mode radiation is trapped at the LFS. The gray area marks the 55--75~GHz frequency range considered here.}
\label{fig,cutoffs}
\end{center}
\end{figure}

\subsection{Ensemble-averaged treatment of reflections}

At frequencies well below the fundamental EC resonance, plasma absorption/emission and the resulting ECE background in both X- and O-mode can be expected to be relatively low. 
Under such optically thin plasma conditions, effects of wall reflections  cannot be ignored, and any antenna will not only see ECE signals emitted along the first pass of its line of sight but in principle from almost anywhere in the plasma. Given that the plasma-facing wall in ITER has a complex 3D structure and composition with reflective properties that may evolve over its lifetime, and that any antenna pattern or receiver view cannot be represented by a single ray, it is not feasible to simulate the effects of wall reflections in every detail. Instead, we take a statistical approach and consider an ensemble average of possible ray paths across the plasma. In practice, this is done by tracing a ray across $N=2000$ wall reflections and using averaged values for this ensemble. This value of $N$ is required to get average values with statistical fluctuations at the 10\% level.

Our approach allows for an arbitrary wall geometry in a poloidal cross section. We have adopted the reference design for the ITER first wall and divertor \cite{fw}, with the vacuum vessel   assumed toroidally symmetric. Figure~\ref{fig,wall} shows this geometry along with the near-perpendicular default viewing geometry {\em A} adopted below. For completeness, we will also consider two alternative sight lines, {\em B}, looking past the inner column, and {\em C}, looking towards the divertor, with viewing geometries summarized in Table~\ref{tab,geom}.

\begin{figure*}
\begin{center}
\includegraphics[width=11cm]{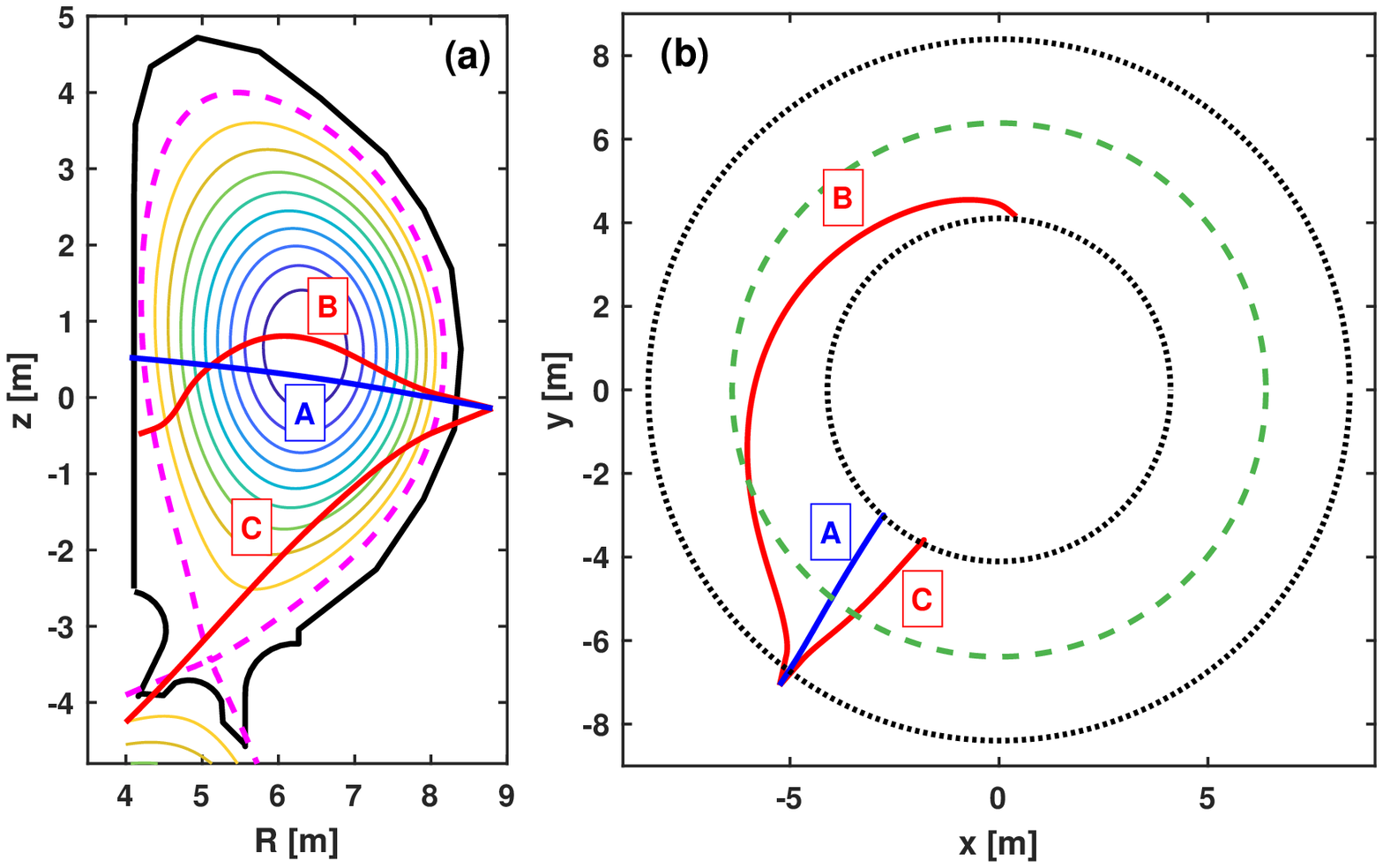}
\caption{Adopted ITER wall geometry in a (a) poloidal and (b) toroidal cross section. The blue line shows our default single-pass receiver viewing geometry {\em A} and red lines the corresponding alternative sight lines {\em B} and {\em C}. Depicted single-pass ray paths result from raytracing at $f=60$~GHz using the baseline plasma scenario; the bending is due to refraction owing to proximity to the X-mode L-cutoff.}
\label{fig,wall}
\end{center}
\end{figure*}

\begin{table}
\caption{\label{tab,geom}Summary of adopted viewing geometries. $\theta$ is the viewing angle relative to horizontal, $\zeta$ is the rotation angle relative to radial  in a toroidal cross section  (counted counterclockwise), and $\phi$ is the resulting viewing angle with respect to the local magnetic field at the assumed detector location of ({\em x},{\em y},{\em z}) = ($5.25$,$-7.07$,$-0.14$)m in the standard ITER global coordinate system used in Figure~\ref{fig,wall}.} 
\begin{indented}
\item[]\begin{tabular}{@{}*{4}{c}}
\br          
Geometry &  $\theta$ & $\zeta$ & $\phi$  \cr
\mr
{\em A}        & $10^\circ$  & $4^\circ$ & $85^\circ$ \cr
{\em B}        & $13^\circ$  & $24^\circ$ & $66^\circ$ \cr
{\em C}        & $-22^\circ$  & $-5^\circ$ & $103^\circ$ \cr
\br
\end{tabular}
\end{indented}
\end{table}

Our treatment of the directional distribution function of reflections off the walls is similar to that of the “diffuse reflection” option of the NOn-Thermal ECE Code (NOTEC) \cite{sill87}. This assumes reflection at an angle drawn from a Gaussian distribution around the angle corresponding to specular reflection, a case we will refer to as randomized reflections and which is illustrated in Figure~\ref{fig,reflection}. Choosing a non-zero width $\sigma$ for this distribution allows for our imperfect knowledge of the detailed wall structure, since we do not include holes, tile gaps, or other oddly angled surfaces. It also enables us to mimic the reflectance properties of the many materials that display a mixture of diffuse and specular reflection (including the extreme case of a fully diffuse Lambertian surface). Here we assume a width $\sigma = 20^\circ$, but as will be shown in Section~\ref{sec,depend}, results are only sensitive to the exact choice of $\sigma$ for cases approaching specular reflection. As the randomization is treated in 3D, randomly reflected rays can leave the plane defined by the incoming and specularly reflected rays.

Defining a ray segment as the ray path between successive reflections, and going backwards  from the receiver, a ray will for each segment $n$ experience a radiation temperature increment $T_{r,n}$. This must reflect $n-1$ times to reach the observer, losing intensity at each reflection due to absorption on the wall and in the plasma. 
The combined multi-pass ray after $N$ reflections will then have radiation temperature
\begin{equation}
T_{rad} = \sum_{n=1}^{n=N}{T_{r,n}} \prod_{i=1}^{i=n-1}{R_{w,i}e^{-\tau_i}}, 
\label{eq,trad1}
\end{equation}
where $R_{w,i}$ and $\tau_i$ are the wall reflectivity and optical depth of the $i$'th segment, respectively.
Adopting now running statistical averages for the single-pass quantities $T_r$, $R_w$, and $\tau$ (i.e.\ setting $\tau = \langle \tau_i \rangle$ etc.), we can discard the segment indices and reduce equation~(\ref{eq,trad1}) to
\begin{equation}
T_{rad} = \sum_{n=1}^{n=N}{T_r(R_w e^{-\tau})^{n-1}}.
\end{equation}
Since $R_w e^{-\tau} < 1$, we can further make use of the geometric series $\sum_{n=0}^{N-1} r^n = (1-r^N)/(1-r)$ to write the radiation temperature as
\begin{equation}
T_{rad} = T_r \frac{1-(R_w e^{-\tau})^N}{1-R_w e^{-\tau}} \simeq \frac{\langle T_r\rangle}{1-\langle R_w\rangle e^{-\langle \tau \rangle}},
\label{eq,trad2}
\end{equation}
where the last expression is valid for a large number of reflections, $N \gg 1$, and in which the final ensemble-averaged quantities are now explicitly indicated with brackets. Note that knowledge of $R_w$ is not needed for computing ensemble averages for $T_r$ and $\tau$, so wall absorption can be included {\em a posteriori}.

\begin{figure}
\begin{center}
\includegraphics[width=7cm]{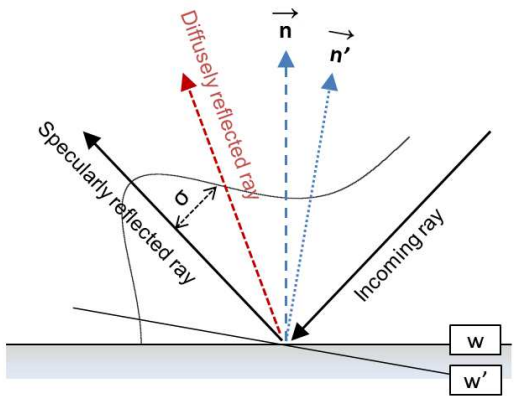}
\caption{2D representation of the reflection geometry for specular and randomized (diffuse) reflections. Specular reflections take place in the plane $w$ of the actual surface with normal vector $n$. Diffuse reflections take place at an angle drawn from a Gaussian distribution of width $\sigma$ around the direction of specular reflection; for the purpose of calculating the associated cross-polarization, these are assumed to take place in the plane $w'$ with normal vector $n'$, determined by the incoming and (randomly drawn) outgoing wave vectors.}
\label{fig,reflection}
\end{center}
\end{figure}

\subsection{Cross-polarization}\label{sec,pol}

Equation~(\ref{eq,trad2}) does not consider the conversion of X- and O-mode waves that may occur as a consequence of wall reflections. To include this, we compute the wave electric field before and after each wall reflection, taking into account the incoming and reflected wave vectors, the magnetic field vector at the reflection point, and the  
 normal vector of the reflecting surface assuming specular reflection. Note that for randomized reflections, this means that the normal
vector will not correspond to the one obtained from assuming toroidal symmetry, but rather to that of the assumed randomly oriented surface as shown in Figure~\ref{fig,reflection}.

To allow for surface roughness, we compute the cross-polarization assuming reflection off a grooved surface, broadly similar to the reflection off “fish-scale like” structures on the first wall considered in \cite{kern96}. This introduces a free parameter, 
namely the phase delay $\Delta \psi$ between the wave components reflected off the bottom and top of the grooves. With our approach of averaging over many randomized reflections, the exact value of this parameter is of little consequence for our ensemble-averaged results for $T_{rad}$.  Hence, we simply take $\Delta \psi = 0$ as our default assumption, corresponding to a smooth (but not necessarily regular) surface, but we discuss the impact of this assumption in Section~\ref{sec,depend}. 

It is intractable to trace every ray when considering reflection-induced cross-polarization, as the number would grow as $2^N$. However, the behaviour of O-mode rays in the high-density plasma scenarios considered here can be used to make a statistical argument. For our frequency range of interest, 
such rays will encounter a cut-off near the plasma edge (at the location shown in Figure~\ref{fig,cutoffs}) and so remain at the edge where the plasma is cold and optically thin, regardless of propagation direction. 
Indeed, raytracing shows that O-mode rays will not pick up any radiation from the plasma, nor get absorbed there, with $T_r$ and $\tau$ staying  $\simeq 0$ for these rays. Instead, an O-mode ray will eventually disappear, each
reflection causing part absorption in the wall and part conversion back to X-mode. To assess the wave power going each way, we first trace an O-mode ray through several hundred reflections. This provides ensemble-averaged X- and O-mode reflectivities 
$R_{OO} + R_{OX} = 1$, where $R_w R_{OO}$ and $R_w R_{OX}$ is the average fraction of an O-mode ray returned in O- and X-mode, respectively, after a {\em single} reflection, and $R_w$ is the mode-independent average wall reflectivity.  With this, we 
can then again use the geometric series to calculate the average fraction $R_{OX}^\ast$ of a reflected O-mode ray converted back into X-mode after $N$ reflections ($\neq R_{OX}$ for $N>1$): 
\begin{equation}
R_{OX}^\ast = R_{OX}\frac{1- (R_{OO}R_w)^N}{1-R_{OO} R_w} \simeq \frac{R_{OX}}{1-R_{OO} R_w},
\end{equation}
where again the last expression applies for $N \gg 1$.

Cross-polarization is then included by assuming that at each reflection a fraction $R_{XX}$ of the reflected ray remains in X-mode, a fraction $R_{XO} R_{OX}^\ast$ is reflected in O-mode but eventually converted back to X-mode, and a fraction $R_{XO}(1-R_{OX}^\ast)$ is converted to O-mode and lost to the wall. Analytically, this is implemented by replacing the wall absorption expressed through $R_w$ in equation~(\ref{eq,trad2}) with $R_{all,X}=R_w(R_{XX}+R_{XO} R_{OX}^\ast)$, so that cross-polarization is effectively considered as an additional absorption mechanism at the wall.
The fraction $R_{XO} R_{OX}^\ast$  is here traced from the same location as $R_{XX}$, so the effects of new launching locations and directions are not included.

The final expression for the X-mode radiation temperature along a receiver line-of-sight then becomes
\begin{equation}
T_{rad} \simeq \frac{\langle T_r\rangle}{1-\langle R_{all,X}\rangle  e^{-\langle \tau \rangle}},
\label{eq,trad3}
\end{equation}
with
\begin{equation}
\langle R_{all,X}\rangle= \langle R_w(R_{XX} + R_{XO} R_{OX}^\ast)\rangle , 
\end{equation}
and
\begin{equation}
R_{OX}^\ast = \frac{\langle R_{OX} \rangle}{1-\langle R_{OO} R_w\rangle},
\end{equation}
where ensemble averages are  indicated with brackets. This expression accounts, in an approximate and statistical manner, for absorption in the plasma and on PFCs, as well as for multiple wall reflections and the associated cross-polarization. Along with the adopted reflection angle distribution function, it also accounts for specular lobes formed by randomized reflection from surfaces that break toroidal symmetry or otherwise do not conform to the simplified wall shape assumed here.

Additional cross-polarization {\em between} wall reflections may be induced by the shear of the magnetic field along the propagation direction and by the Faraday effect. However, the general condition for the waves to remain a local characteristic mode in the presence of magnetic shear, equation~(12) of \cite{segr90}, is easily met for nearly all our X-mode ray segments, except for a few cases that correspond to entirely negligible $T_{rad}$ increments. In addition, the average Faraday rotation angle of our X-mode segments is small, $\sim 4^{\circ}$. Within our 55--75~GHz range, any O-mode component is evanescent inside the ITER density pedestal and encounters a cutoff outside it, so no significant O-- to X-mode conversion takes place outside wall reflections. The net impact of the Faraday effect and the magnetic shear is thus limited to a small net conversion of X-- to O-mode. We assume this effect to be negligible compared to the 20--30\% average polarization conversion taking place at the wall (see Section~\ref{sec,depend}), and the associated reduction in X-mode intensity to be within our statistical uncertainties on $T_{rad}$.

\subsection{Wall reflectivity}

The use of equation~(\ref{eq,trad3}) requires knowledge of the wall reflectivity $R_w$ for microwaves. Determination of this parameter for the ITER first wall will be a goal of the ITER ECE diagnostic itself \cite{udin12}, but for now we must rely on an estimate. For this, we note that
the fraction of absorbed wave power per reflection at normal incidence is 
\begin{equation}
P_{\rm abs} = 4 \xi  (\pi f \epsilon_0 \rho)^{1/2}, 
\label{eq,reflec}
\end{equation}
where $f$ is the wave frequency, $\epsilon_0$ the vacuum permittivity, $\rho(T)$ is the temperature-dependent electrical resistivity of the relevant PFC material, and $\xi$  is an empirical factor accounting for possible surface roughness. The latter is often taken to be $\xi \approx 2$ (e.g., \cite{siri15}), but plasma coating may increase this to $\approx 3$ \cite{kasp12}.
Taking $\xi = 2$--3, $T=300^\circ$~C for the vessel wall \cite{raff14}, and $T=1100^\circ$~C for the divertor \cite{hira16} would suggest
$P_{\rm abs} = 0.4$--0.6\% for the ITER Be walls, 
0.6--0.9\% for the W divertor, and 1.0--1.5\% for any stainless-steel surfaces at $f = 60$~GHz, with a frequency dependence that can be ignored here. 

Hence, for plain surfaces, the typical wall reflectivity will not exceed $R_w = 1 - P_{\rm abs} \approx 0.995$, and this is made more relevant still by the consideration that the adopted value represents an average over all incidence angles (see below). Furthermore, the presence of, e.g., tile gaps, holes, ports, and antennas in the ITER wall that can act as microwave sinks will reduce the {\em effective} $R_w$ well below this value. Indeed, \cite{denk18} lists an effective value of $R_w \approx 0.9$ for the wall in ASDEX Upgrade (subject to the assumptions
underlying their ECE modelling), while \cite{born83,fido96} quote typical values of $R_w = 0.7$--0.8 for tokamak metallic walls in general. However, values down to $R_w = 0.6$--0.7 are sometimes assumed for the ITER walls \cite{alba05,mina15} at comparable frequencies. This is similar to empirical values for carbon/graphite walls in, e.g., JET \cite{barr10} and DIII-D \cite{aust97} and to assumed values for such walls in existing models of generic \cite{kern96} and specific (TFTR; \cite{born96}) fusion devices. 
It should be mentioned that such empirical, ECE-based estimates of $R_w$ are generally model-dependent and hence do not necessarily transfer directly to the model presented here. We therefore consider a wide range of possible $R_w$, taking $R_w = 0.9$ as our default value and plausible upper limit while assessing the sensitivity of our results to reflectivities down to $R_w=0.6$.

Our approach does not account for any dependence of wall reflectivity on incident angle. At shorter wavelengths, the reflected intensity $I'$ for a non-Lambertian (metallic) surface may be empirically modelled as $I' \propto R_w \mbox{cos}^\eta \psi$, or as sum of such terms in the case of multiple reflection lobes  \cite{reic09,aume12,koca16}. Here $R_w$ is the surface reflectivity at normal incidence, $\psi$ is the angle between the viewing line and the surface normal, and $\eta$ depends on the surface structure and wavelength. Being unaware of relevant available data for the ITER PFCs that constrain any dependence on $\psi$ in the microwave range, we have refrained from incorporating such a dependence here. Since our approach considers only ensemble averages of a given ray, it should anyway not be necessary to account for this in detail, as only the average value $\langle R_w\rangle$ assumed for the wall reflectivity is relevant.

\section{Results}\label{sec,results}

Before describing the ECE spectra resulting from our simulations, we first highlight a few features of the underlying raytracing in the adopted geometry.
Here it is instructive to compare the case of  specular reflection, i.e.\ with an angle distribution function with $\sigma=0^\circ$, to our default case of randomized reflections with $\sigma=20^\circ$. Figure~\ref{fig,raypaths2} shows X-mode ray paths at 60~GHz in these two cases for the first 50 reflections from the antenna, highlighting two points of importance for the discussion to follow. 
\begin{figure}
\begin{center}
\includegraphics[width=8.2cm]{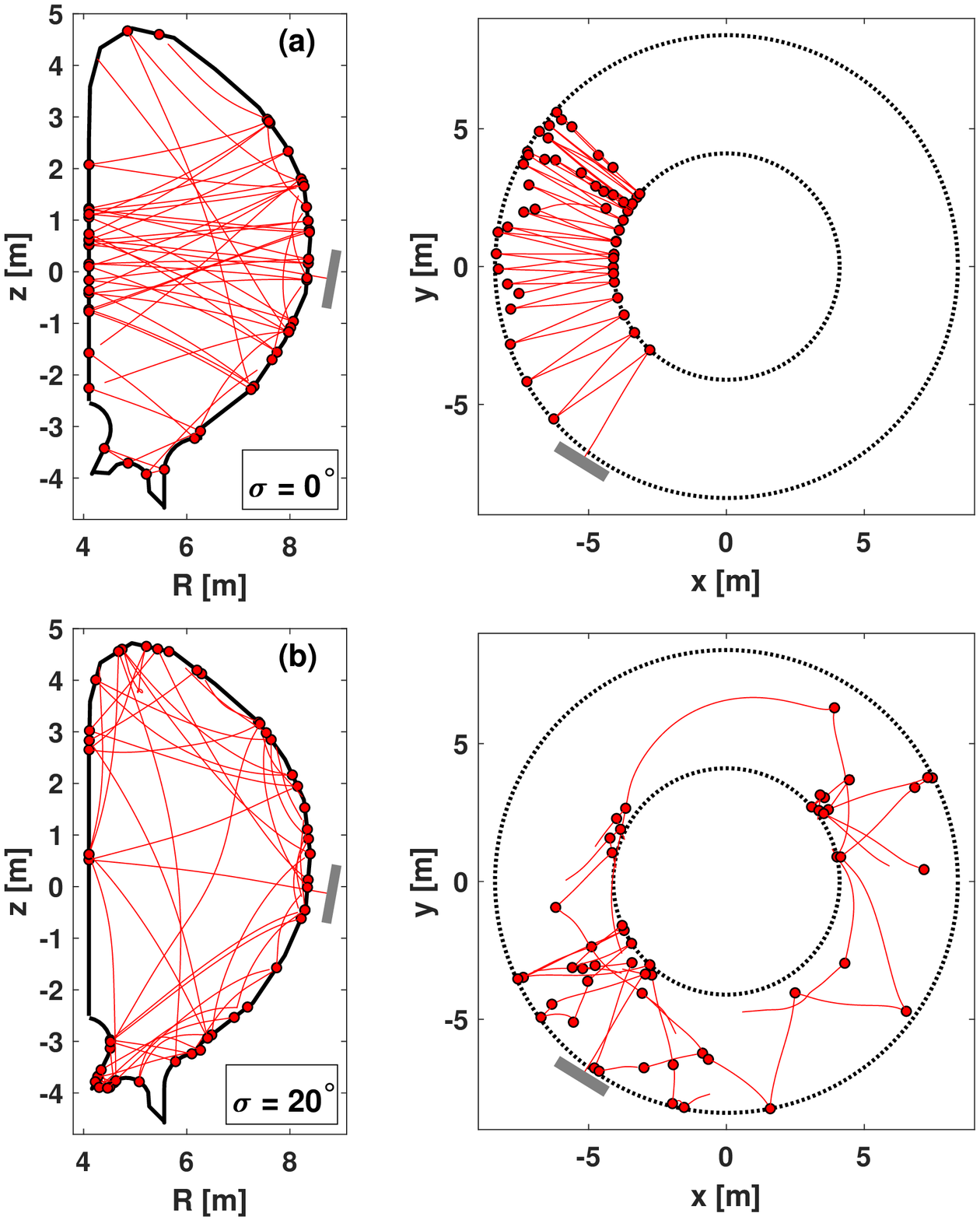}

\vspace{3mm}

\includegraphics[width=8.2cm]{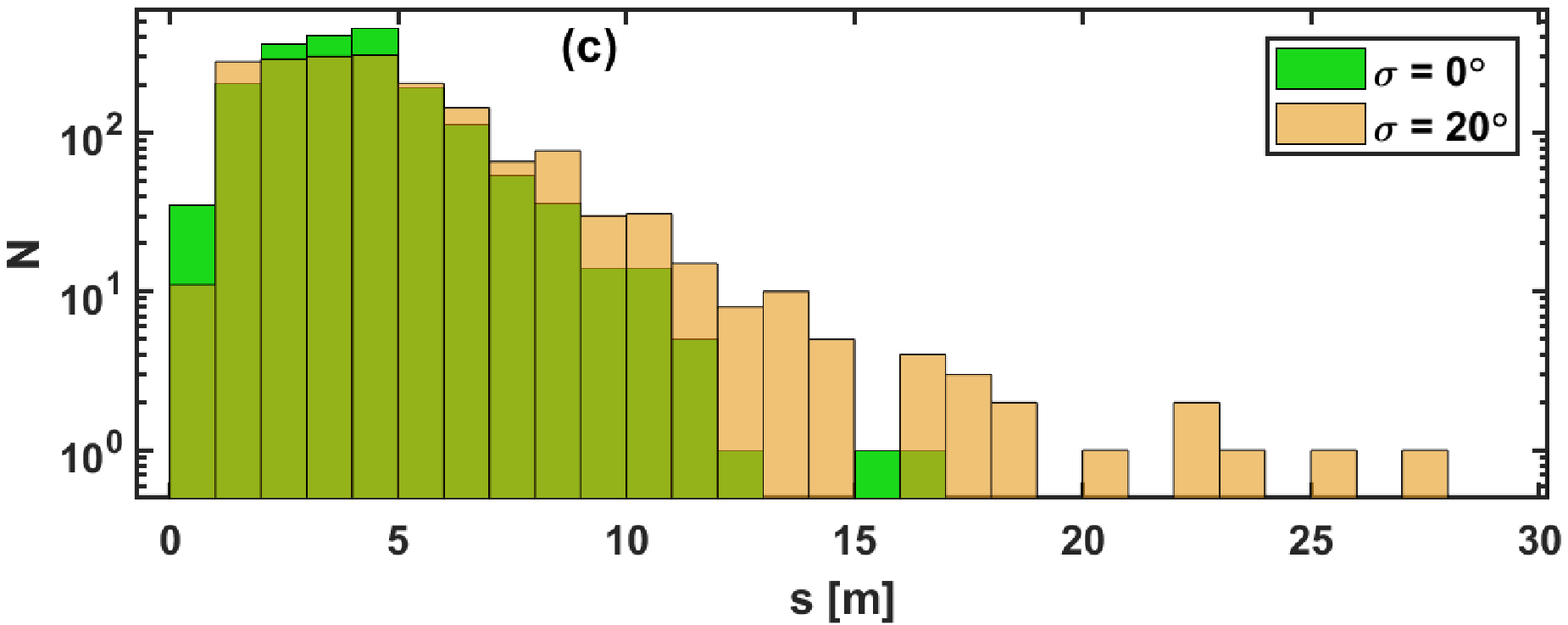}
\caption{Example X-mode ray paths for the first 50 reflections going backwards from the antenna for viewing geometry {\em A} in the ITER baseline plasma scenario in the case of (a) specular reflection, (b) our default case with $\sigma=20^\circ$. Filled circles mark the associated reflection points and gray 
rectangles the assumed antenna location and orientation. (c) Example distributions of the length $s$ of individual ray segments for all $N=2000$ reflections in both cases. Randomized reflections yield a broader distribution with a tail towards large $s$.}
\label{fig,raypaths2}
\end{center}
\end{figure}

First, for a completely smooth, toroidally symmetric wall (corresponding to $\sigma=0^\circ$), rays in our near-perpendicular viewing geometry tend to stay close to radial in a toroidal cross section (Figure~\ref{fig,raypaths2}a). In this case, a typical ray segment will probe the hot, dense plasma core only over a modest path length before losing power in wall reflections.
Doppler broadening of the EC resonance is expected to be limited for these radial segments. In contrast, diffuse reflections can generate highly oblique ray segments with significant Doppler shifts and long path lengths through the plasma center between reflections (Figure~\ref{fig,raypaths2}b and c).
Second, rays can become temporarily trapped in the divertor region (Figure~\ref{fig,raypaths2}b), undergoing multiple reflections here before again proceeding into the plasma core. We find that, on average, 20--25\% of all reflections take place in the region $z < -3$~m (where $z$ is the vertical distance from the center of
the tokamak, cf.\ Figures~\ref{fig,wall} and \ref{fig,raypaths2}). In this fashion the divertor partially acts like a beam dump.

\subsection{Dependence on frequency and plasma conditions}\label{sec,resfreq}

For our default assumptions, i.e.\ viewing geometry {\em A} as shown in Figure~\ref{fig,wall} and an angle distribution function with $\sigma=20^\circ$,
Figure~\ref{fig,raypaths} shows the resulting radiation temperatures in the ITER baseline plasma scenario for the first $N=1000$ reflections from the antenna. Results are shown both for the individual segments of one particular (randomized) ray path with no integration along the path,
and for a ray based on ensemble-averaged values according to equation~(\ref{eq,trad3}).

From this, we note that the contributions from individual ray segments typically consist of stretches with only negligible $T_{rad}$
increments (e.g., when the ray travels along the plasma edge or becomes temporarily trapped in the divertor region), interspersed with consecutive but relatively short
sections with larger contributions (e.g., when the ray traverses the plasma center with a large toroidal component of its wave vector). Examples of both types of ray segments are
visible in Figure~\ref{fig,raypaths2}; note that
tracing rays through many reflections is important for describing the statistical probability of such events, which cannot be captured
by simply assuming random launching directions and locations.
The radiation temperature of the ensemble-averaged ray is seen to build up over a few tens to a few hundred reflections, saturating once the absorbed energy per reflection equals the average $T_{rad}$ increment.
Incorporating a large number of reflections in the ensemble averaging is thus not required for $T_{rad}$ to converge, only for reducing the statistical error on the result.

\begin{figure}
\begin{center}
\includegraphics[width=8.3cm]{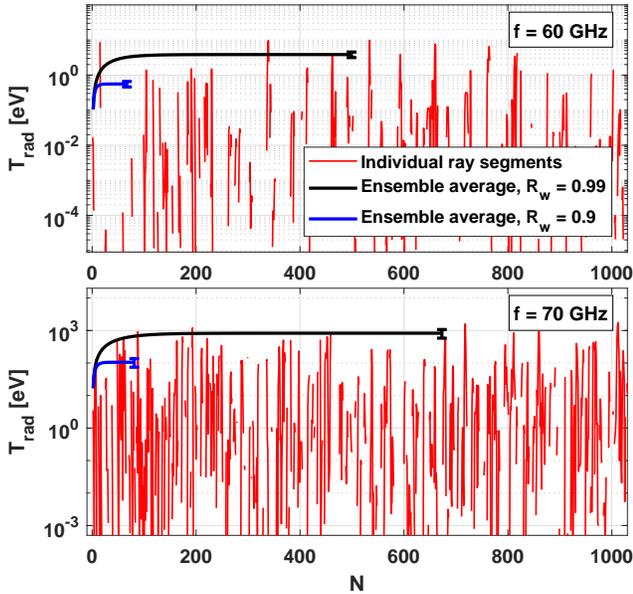}
\caption{Examples of the contribution to the radiation temperature for individual ray segments as a function of the number of reflections taken. Also shown is the ensemble-averaged result for two different values of the wall reflectivity $R_w$, our default value of $R_w=0.9$ and, for comparison, an extreme case with $R_w=0.99$. 
Results apply for the ITER baseline scenario. Error bars show the final statistical errors on $T_{rad}$ of the averaged rays.}
\label{fig,raypaths}
\end{center}
\end{figure}

Figure~\ref{fig,Trad1}a shows results of combining such calculations into ECE spectra for a range of assumed wall reflectivities $R_w$ in the ITER baseline scenario.
The typical statistical uncertainty (1$\sigma$ error on the mean) on the ensemble-averaged $T_{rad}$ shown here is $\sim$\,10\%. The observed ECE signal can be identified with the emission associated with 
the fundamental resonance, relativistically downshifted and broadened to frequencies below the X-mode R-cutoff at the outboard plasma edge \cite{born96} of $f \approx 116$~GHz (corresponding to $\omega \lesssim 0.78\omega_c$ for this plasma scenario), see also Figure~\ref{fig,cutoffs}. 
%Raytracing shows that most of the observed radiation at 60~GHz originates across a broad region extending from $R \approx 6.0-7.5$~m. 
As expected for frequencies around $\sim 0.5 \omega_c$ \cite{born96}, $T_{rad}$ is found to rise very rapidly with frequency for the assumed plasma conditions, increasing 
by five orders of magnitude across our considered range. A  dependence on wall reflectivity is also seen, but  $T_{rad}$ remains below 100~eV 
at frequencies $f \lesssim 70$~GHz even for high $R_w$. An analytical relation that matches all numerical predictions shown in Figure~\ref{fig,Trad1}a to within our uncertainties
is provided by
\begin{equation}
\mbox{log}_{10}(T_{rad})  =1.24(f-54.3)^{0.545} + R_w - 4.47, 
\label{eq,analyt1}
\end{equation}
with $T_{rad}$ in eV, $f \in [55; 75]$ in GHz, and $R_w \in [0.6; 0.9]$.

\begin{figure}
\begin{center}
\includegraphics[width=8.2cm]{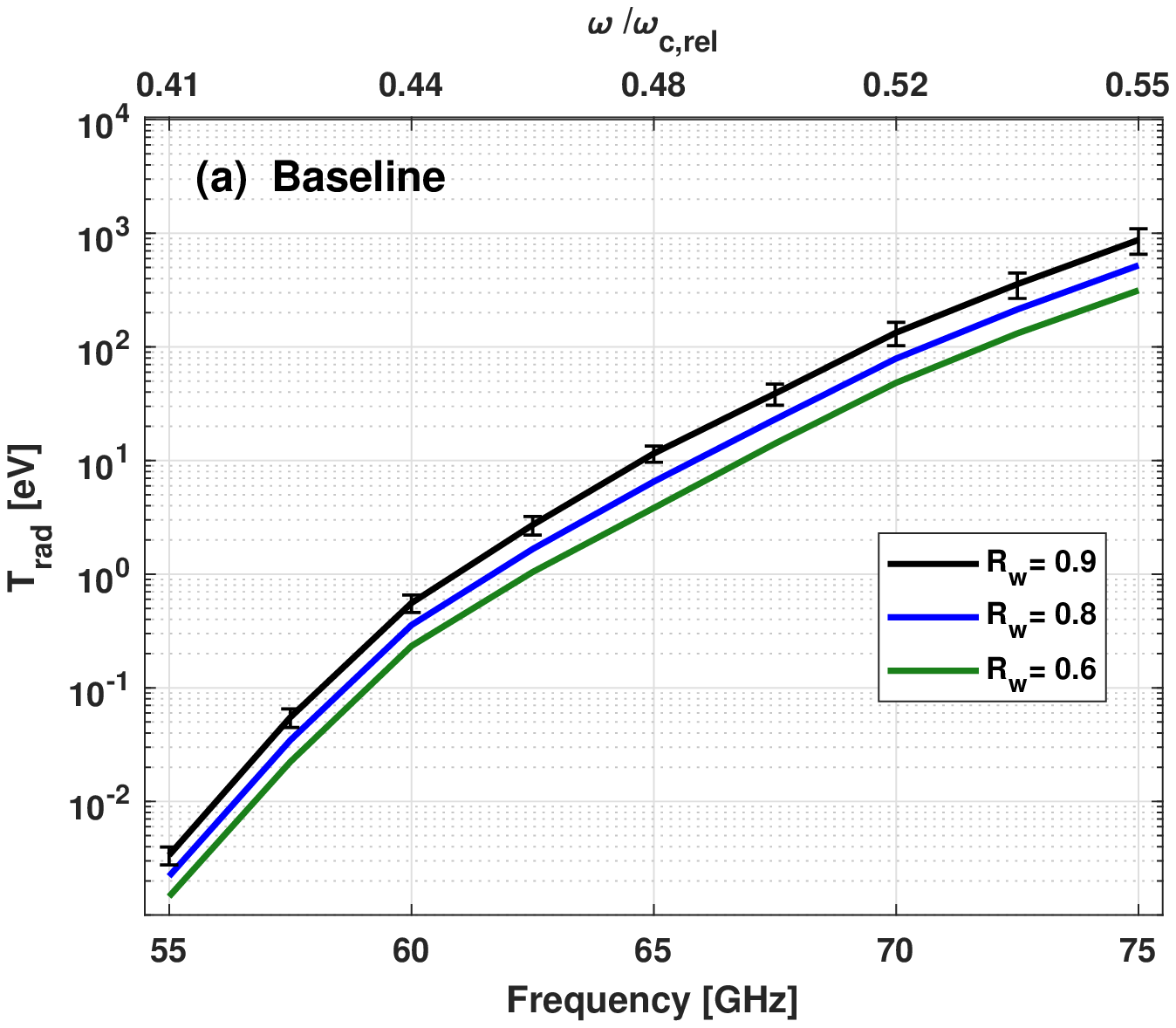}\vspace{2mm}

\includegraphics[width=8.2cm]{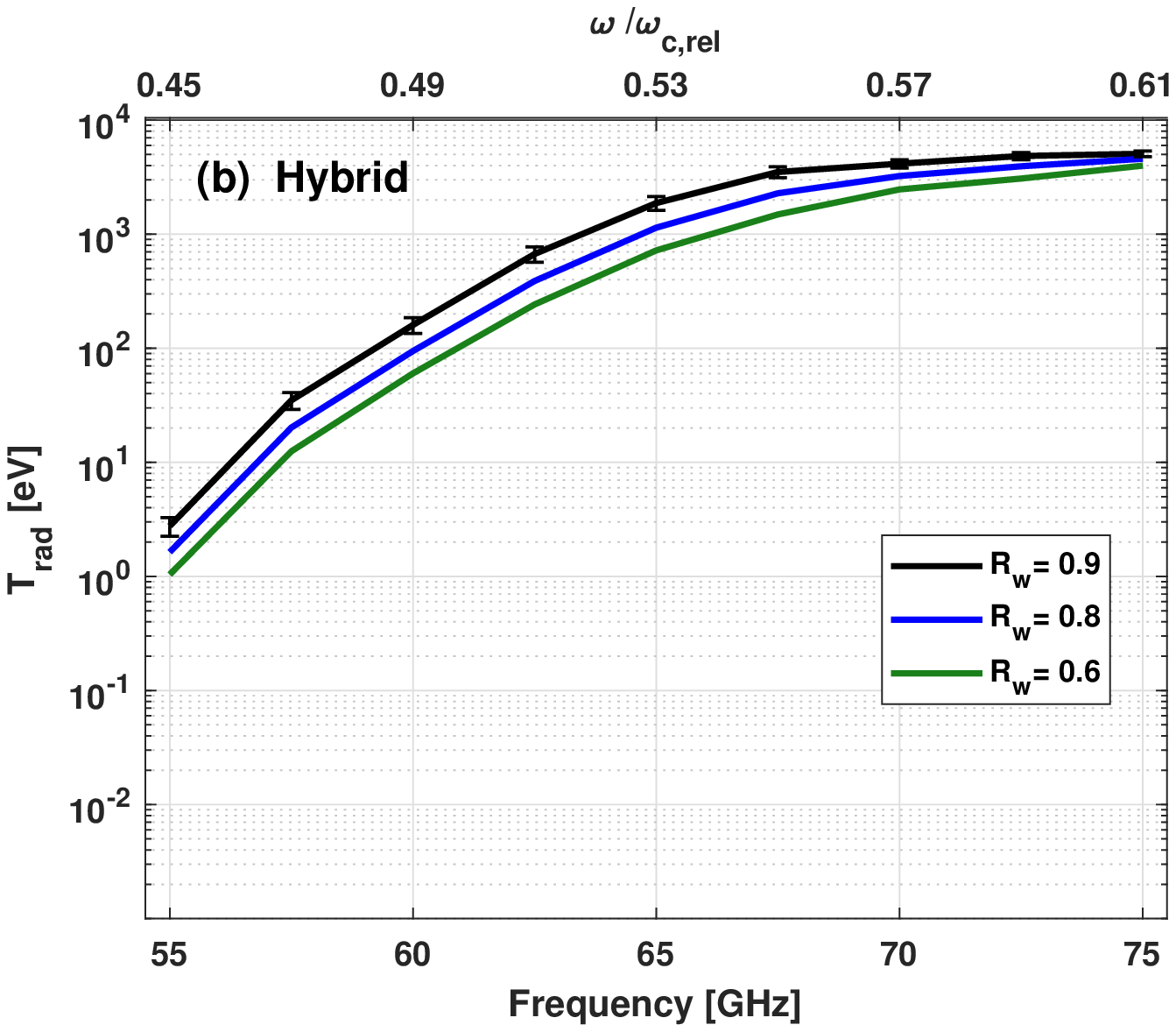}
\caption{Simulated X-mode ECE spectra for (a) the ITER baseline scenario and (b) the hybrid scenario for different wall reflectivities $R_w$. Representative statistical uncertainties are shown for the case $R_w = 0.9$. The same vertical axis is used in both plots, and the top horizontal axis gives the frequency in terms of the relativistically downshifted first harmonic at the magnetic axis, $\omega_{c,rel}$.}
\label{fig,Trad1}
\end{center}
\end{figure}

Corresponding results for the ITER hybrid scenario also introduced in Section~\ref{sec,ray} are shown in Figure~\ref{fig,Trad1}b. At the higher $T_e$ encountered here, the predicted $f \lesssim 70$~GHz ECE level becomes substantial, two orders of magnitude above that of the baseline scenario. 
Radiation temperatures at the keV-level are here predicted to persist below the 70~GHz lower frequency limit of the planned ITER ECE diagnostic and, for high $R_w$, into the $60\pm 5$~GHz frequency range to be covered by the CTS receivers.
We also note how the high-frequency results in this case become less sensitive to the assumed value of $R_w$. This is a consequence of plasma absorption beginning to dominate wall absorption, due to increasing relativistic downshift and broadening of the fundamental resonance with increasing $T_e$. 

At the frequencies considered here, the ones further from the cold fundamental resonance and closer to the corresponding X-mode L-cutoff carry a
 very strong dependence of $T_{rad}$ on electron temperature. For example, at $f = 60$~GHz, Figure~\ref{fig,Trad1} suggests an increase in $T_{rad}$ of a factor $\sim 300$ for a modest 20\% increase in core $T_e$ between the two plasma scenarios. This is not only caused by the higher $T_e$ itself in the hybrid scenario, but also by the lower density and the associated reduction in refraction due to the resulting downshift of the X-mode L-cutoff. As the optically thin ECE is determined by the condition that the average $T_{rad}$ increment equals the energy absorbed in reflections, it is dominated by contributions from oblique ray segments subject to strong Doppler broadening of the EC resonance and with long path lengths through the plasma core between reflections. At higher $T_e$ and lower $n_e$, the reduced refraction implies that rays can more easily propagate into -- and within -- the plasma core, and
this is particular true for the long oblique ray segments that dominate the contribution to $T_{rad}$. Indeed, we find that the average ray segment is $\sim$\,25\% longer in the hybrid scenario than in the baseline one, with a tail towards large segment lengths, reminiscent of the case of specular vs.\ randomized reflections illustrated in Figure~\ref{fig,raypaths2}c.

The temperature dependence on $T_{rad}$ at fixed $n_{e,0}$ is explored in more detail  in Figure~\ref{fig,Tscaling} which illustrates the variation in predicted $T_{rad}$ at a specific frequency when scaling the temperature profile of the ITER baseline scenario (the relativistic O-mode cutoff remains at the plasma edge at all $T_e$ for these densities, so equation~\ref{eq,trad3} remains valid). The considered $T_e$ range covers the lowest $T_{e,0} \approx 15$~keV for which the {\em Warmray} integration routine yields finite $T_{rad}$ for the baseline scenario, to the maximum $T_e \sim 40$~keV at which ITER is potentially capable of operating. 
While the ECE is predicted to reach keV-levels at 70~GHz for high wall reflectivities in both the baseline and hybrid plasma scenarios, Figure~\ref{fig,Tscaling}
confirms the strong sensitivity of our results to assumptions regarding plasma conditions at the lower end of our frequency range. 
An ITER scenario with lower core $T_e$ but similar temperature profile peaking as assumed here for the baseline scenario would imply a dramatically lower ECE signal at 60~GHz.  
In contrast, for hotter plasmas, the predicted ECE spectrum becomes increasingly flat across our frequency range. 

\begin{figure}
\begin{center}
\includegraphics[width=8.2cm]{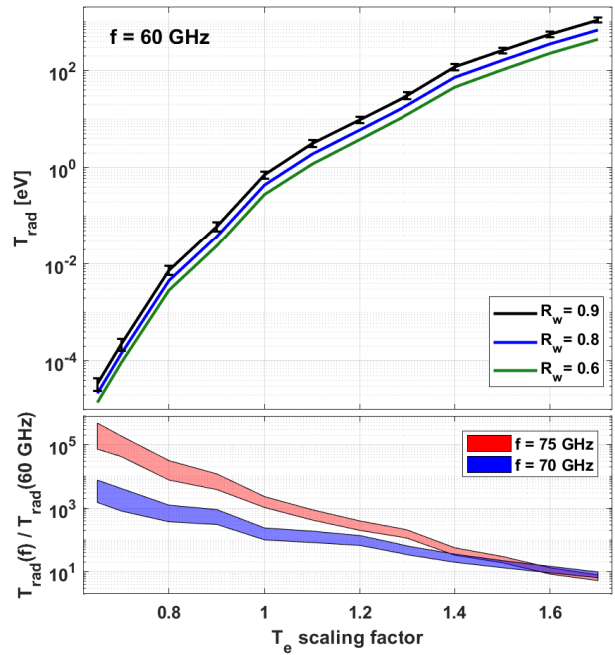}
\caption{Dependence of $T_{rad}$ at 60 GHz on scaled electron temperature of the ITER baseline scenario. Representative uncertainties are shown for a wall reflectivity $R_w=0.9$. Bottom panel shows the ratio of $T_{rad}$ at 75 and 70~GHz, respectively, to that at 60~GHz for $R_w = 0.9$. The horizontal axis range corresponds to a core $T_e$ ranging from 15--40~keV.}
\label{fig,Tscaling}
\end{center}
\end{figure}

Figure~\ref{fig,Tscaling} demonstrates that the X-mode ECE at frequencies below the 70~GHz lower limit of the planned ECE instrumentation remains a powerful diagnostic of plasma temperature in ITER, regardless of the detailed properties of the plasma-facing components. We also note that the planned 55--65~GHz frequency range of the ITER CTS system was chosen in part to minimize the expected ECE background, but that this is associated with increased sensitivity to refraction. Any CTS diagnostic in a DEMO-like device (for which $T_{e,0} \sim 30$--40~keV) might thus benefit from operating closer to $\omega_c$, as this could reduce effects of refraction without necessarily compromising the anticipated signal-to-noise ratio due to increased ECE background.

Based on results similar to those shown in Figure~\ref{fig,Tscaling}, we constructed approximate analytical relations that determine the frequency $f'$ above which the predicted  $T_{rad}$ exceeds specific levels in the baseline plasma scenario as a function of the assumed $T_{e,0}$. The point at which the ECE radiation temperature may be considered significant will necessarily depend on context, so here we provide relations that give the maximum $f'$ at which $T_{rad}$ remains below 100~eV,
\begin{equation}
f' (T_{rad} < 100\mbox{ eV}) = -1.04T_{e,0} -5.2 R_w^{4.7} + 98.2,
\label{eq,analyt2}
\end{equation}
as well as the value of $f'$ for which $T_{rad}$ remains below 1\% of $T_{e,0}$,
\begin{equation}
f' (T_{rad} < 0.01T_{e,0})= -0.88 T_{e,0} -4.6 R_w^{4.7} + 96.1,
\label{eq,analyt3}
\end{equation}
valid for $f' \in [55; 75]$ in GHz, $T_{e,0} \in [15; 40]$ in keV, and $R_w \in [0.6; 0.9]$. Either of these two limits on $T_{rad}$ should be easily detectable by any sensitive radiometer.

\subsection{Impact of model assumptions and cross-polarization}\label{sec,depend}

For a given plasma scenario, the salient free parameters in our model other than $R_w$ are, in practice, the antenna orientation and the width $\sigma$ of the Gaussian distribution function for reflection angles. For the latter, recall that $\sigma = 0^\circ$ corresponds to specular reflection, whereas large values imply diffuse (Lambertian) reflection. To expose the effect of varying these, Figure~\ref{fig,angle} shows the predicted radiation temperature for the ITER baseline scenario obtained for different $\sigma$ and viewing geometries. These results apply at a specific frequency and wall reflectivity, as varying these parameters would simply scale each curve according to Figure~\ref{fig,Trad1} without 
introducing further dependencies on $\sigma$.

\begin{figure}
\begin{center}
\includegraphics[width=8.2cm]{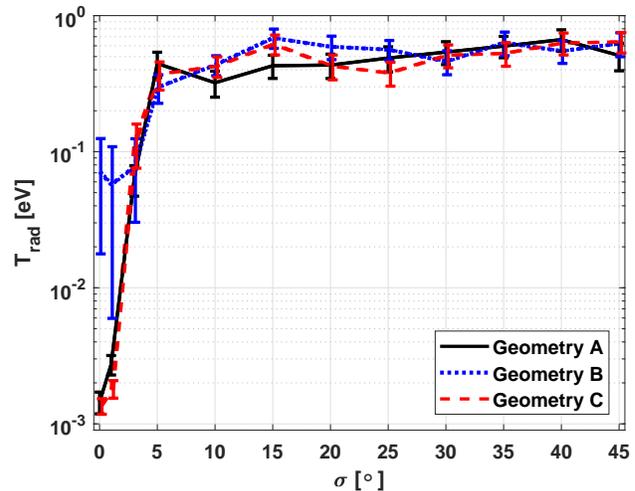}
\caption{Variation of $T_{rad}$ as a function of the assumed width $\sigma$ of the angle distribution function for the ITER baseline scenario, computed for $f=60$~GHz and 
$R_w = 0.9$. The curves show the impact of different antenna orientations as indicated in Figure~\ref{fig,wall}.}
\label{fig,angle}
\end{center}
\end{figure}

The most prominent feature is the strong dependence on $\sigma$ at low values of this parameter, with specular reflection implying significantly lower ECE levels than when allowing for reflection lobes. As mentioned above, the $T_{rad}$ increments are dominated by contributions from oblique ray segments with long path lengths through the plasma core and significant
Doppler shifts, but specular reflections that generate such ray segments are not favored by the assumed wall geometry. This differs from the case of diffuse reflections, which, aided by refraction, can produce much longer and more oblique ray segments and hence higher $T_{rad}$  (cf.\ again Figure~\ref{fig,raypaths2}).

As expected, specular reflection also introduces a dependence on the viewing geometry, as the single-pass ray path traced from the antenna becomes more important.
As an example, the longer and much more oblique first-pass sightline in geometry {\em B} contributes to a significantly higher $T_{rad}$ at small $\sigma$ than in our default geometry, albeit with a large statistical uncertainty induced by the resulting significant variation among the individual ray segments. In contrast, the rays in geometry~A remain closer to radial at low $\sigma$ and $T_{rad}$ stays comparatively low. For similar reasons, geometry {\em B} with a viewing angle $\phi=66^\circ$  with respect to the magnetic field also results in higher $T_{rad}$ than geometry {\em C}  ($\phi=110^\circ$), a difference which is unrelated to whether the viewing direction has a component parallel or anti-parallel to the magnetic field. In other cases, specularly reflected X-mode waves can yield higher $T_{rad}$ for $\phi>90^\circ$ than at  $\phi<90^\circ$, in contrast to the situation here, for example for a plasma containing a supra-thermal population of electrons drifting along the magnetic field relative to the bulk electrons  \cite{born96}.

For diffuse reflections, the direction of the first pass is seen to be of little consequence, so a receiver will observe a largely uniform ECE background across the plasma irrespective of its viewing geometry. Hence, the exact choice of $\sigma > 10^\circ$ does not impact significantly on predicted ECE levels, and our use of $\sigma  = 20^\circ$ for all calculations above should thus not introduce appreciable bias. This is encouraging, given that the material properties of all plasma-facing components in ITER are difficult to specify in full detail at this stage and may furthermore evolve during ITER operation. 

The observed ECE levels for X-mode waves can be significantly reduced by cross-polarization. In its absence, rays with large toroidal components can,
as mentioned, potentially build up very large radiation temperatures, easily reaching the keV range in the frequency range considered. However, such toroidal rays will typically reflect at large angles relative to the surface normal and hence undergo significant cross-polarization (leading to large $R_{XO}$ and small $R_{XX} = 1-R_{XO}$). The subsequent absorption of a significant part of the O-mode component on the wall strongly inhibits the build-up of the radiation temperature. 
For low $\sigma$, such toroidal ray segments are rare and cross-polarization of X-mode waves remains small, but increasingly diffuse reflections will enhance the power transfer from X- to O-mode. 

This is illustrated  in Figure~\ref{fig,pol}a, which shows ensemble-averaged values of the different reflection coefficients described in Section~\ref{sec,pol} as a function of 
$\sigma$. We also plot the X-mode conversion coefficient $p_x$, i.e.\ the fraction of power transferred from X-mode to O-mode in each wall reflection. In our ensemble-averaged approach this is given by $\langle p_x\rangle = \langle R_{XO}/R_{XX}\rangle = 1/\langle R_{XX}\rangle -1$.
As is evident, $\langle p_x\rangle$ generally increases with $\sigma$ but remains independent of viewing geometry. 
For completeness, we further show the variation of all reflection coefficients with $R_w$ in Figure~\ref{fig,pol}b. This illustrates the modest deviation of $R_{all,X}$ from $R_w$, and how all other coefficients vary only slowly, if at all, with $R_w$ and with no systematic dependence on frequency. 

\begin{figure}
\begin{center}
\includegraphics[width=8.3cm]{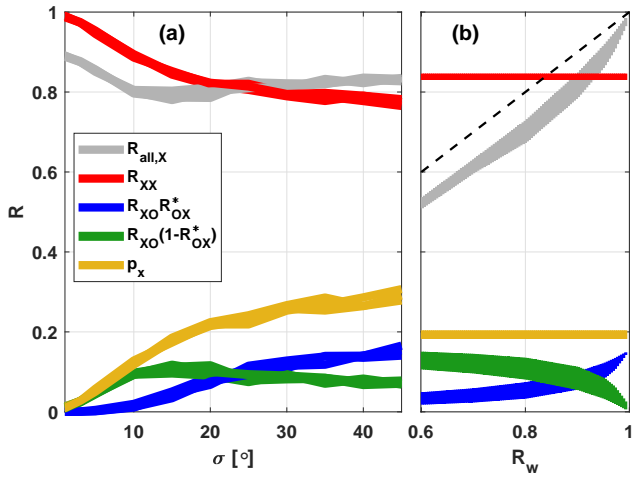}
\caption{Wall reflection coefficients including mode conversion as a function of (a) $\sigma$ at fixed $R_w = 0.9$, and (b) $R_w$ at fixed $\sigma = 20^\circ$. Both plots are for the
ITER baseline scenario at $f=60$~GHz. The width of each curve in (a) represents the variation among the adopted viewing geometries ({\em A}--{\em C}), and in (b) the statistical variation across the considered frequency range for geometry~{\em A}. Dashed curve in (b) shows the case $R=R_w$.}
\label{fig,pol}
\end{center}
\end{figure}

For our default model assumptions, we find $\langle p_x\rangle$ to be in the range  0.2--0.3 for our two plasma scenarios.
This is comparable to typical values assumed elsewhere \cite{alba05} and found experimentally in existing devices \cite{aust97,barr10}. Here it must be emphasized that we have assumed reflection off surfaces that may be irregular but are smooth on scales corresponding to the considered wavelength, such 
that the incident waves are assumed to be reflected uniformly with no phase shifts $\Delta \psi$ introduced across the reflected wave front. If allowing for non-zero $\Delta \psi$, for example as a means of mimicking the presence of grooved surfaces and tile gaps, then the average $\langle p_x\rangle$ increases somewhat at fixed $\sigma$, as seen in Figure~\ref{fig,dpsi}. However, the effective X-mode wall reflection coefficient $\langle R_{all,X}\rangle$ remains nearly constant, and consequently, this has no significant or systematic impact on the simulated radiation temperature.
\begin{figure}
\begin{center}
\includegraphics[width=8.8cm]{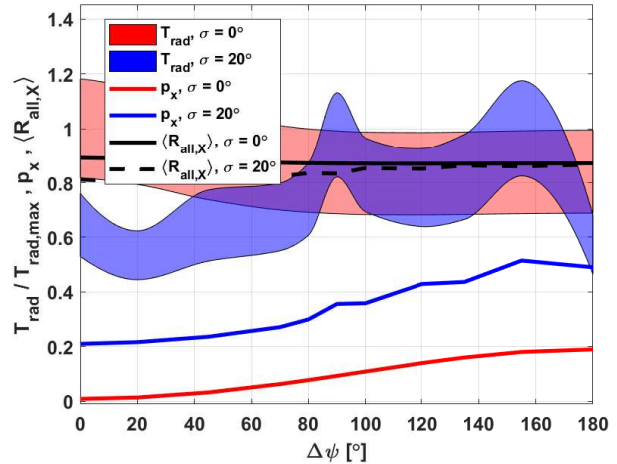}
\caption{Radiation temperature normalized to its maximum value (shaded curves indicating the $\pm 1\sigma$ error interval) and mode conversion coefficient $p_x$ as a function of the maximum phase shift $\Delta \psi$ of reflected waves induced by a grooved surface. Results are shown for the baseline scenario at 60~GHz and $R_w=0.9$, assuming both specular reflection ($\sigma = 0^\circ$) and our default case ($\sigma = 20^\circ$).}
\label{fig,dpsi}
\end{center}
\end{figure}

\subsection{Comparison to measurements and benchmarking with the SPECE code}

The pronounced frequency dependence predicted here at the relevant $\omega < \omega_c$ mirrors earlier results obtained for rather different plasma conditions \cite{born96} or model assumptions \cite{fari08}. The agreement between predicted and measured mode conversion coefficients $\langle p_x\rangle$ is also encouraging, but it would nevertheless be desirable to benchmark our predictions against experiments and/or other codes suitable for predicting ECE levels in ITER at $\omega < \omega_c$. 

While we have been unable to identify existing experimental results that can be directly compared to our predictions, we note for completeness that
X-mode ECE levels of 20--25~eV were measured in TFTR at 60~GHz \cite{rhee92,mach95,mach97} during various hot-ion "supershots". This includes discharge \#55851 with $B_t = 5.1$~T,
$T_{e,0} = 12$~keV, $T_{i,0} = 28$~keV, $n_{e,0} =  9 \times 10^{19}$~m$^{-3}$, and auxiliary heating of $P_{NBI} = 25$~MW \cite{budn92,budn18}. At face value, our predictions for ITER would suggest a lower ECE level at the relevant $T_{e,0}$. 
However, in such TFTR plasmas with strong external heating and fusion-born ions, the likely presence of a non-thermal population of relativistic electrons could contribute significantly to the subharmonic X-mode ECE signal, which might then be expected to exceed our nominal prediction.  
Differences in electron profiles, machine size, wall geometry, magnetic field structure, auxiliary heating power density, as well as the absence of a divertor in TFTR (which can act as an ECE beam dump, as explained in Section~\ref{sec,results}) also make a direct comparison far from straightforward.

Given the scarcity of experimental results that can be directly compared to our predictions, we have instead benchmarked our results against those of the SPECE code \cite{fari08} as an initial means of validation. SPECE was developed for use at JET but has also been employed for ECE predictions at ITER at $\omega < \omega_c$ \cite{fari08}. The code solves equation~(\ref{eq,trad}) using the fully relativistic dispersion relation
 and accounts for the antenna pattern by simulating multiple rays. We have here represented this pattern by a Gaussian beam with a waist of 5~cm at the receiver, the mean value for the seven ITER CTS receivers in the present diagnostic design. Like {\em Warmray}, SPECE takes as input electron profiles, magnetic equilibrium, and receiver location and orientation, and wall absorption and cross-polarization are assumed to occur after each pass across the plasma volume. However, SPECE does not self-consistently compute the mode conversion coefficient $p_x$, so this must be supplied for multi-pass rays along with the assumed $R_w$ as empirical input parameters. 
Here we have adopted $p_x=0.32$, consistent with values obtained from our {\em Warmray} simulations (Section~\ref{sec,depend}), along with $R_w=0.65$, in broad agreement with values from JET \cite{barr10}. 

SPECE further assumes specular reflection between parallel walls, and estimates the multi-pass emission in the limit of infinite wall reflections by computing only the last path leading to the receiving antenna, assuming that all previous passes occur along equivalent paths. The optical depth is then the same for all  passes, and the radiation temperature is computed for odd and even passes by integrating along the line of sight towards and away from the antenna. 
For a straightforward comparison between the two codes, we have therefore assumed our default near-perpendicular viewing geometry, along with  $\sigma=0^\circ$ for the angle distribution function. 

Results for otherwise identical plasma conditions (our adopted ITER baseline scenario) are shown in Figure~\ref{fig,spece}. 
Overall, the agreement between the two predictions is excellent, considering
the large dynamical range in $T_{\rm rad}$. Both the emission and absorption coefficients and hence the resulting radiation temperatures and optical depths are well matched between the two codes in terms of overall scaling and the strong frequency dependence. 
The only possible exception is at the lowest frequencies considered, where our method suggests somewhat lower $T_{rad}$, although with large errors. This may be attributed to numerical artefacts related to the effective resolution of our algorithm, as the {\em Warmray} integration routine yields $T_{rad} \equiv 0$ for levels $\lesssim 10^{-6}$~eV (which is several orders of magnitude below what can be resolved experimentally). The good agreement seen in Figure~\ref{fig,spece} further demonstrates that our use of a weakly relativistic absorption coefficient in equation~(\ref{eq,trad}) is  sufficient for the plasma conditions considered here. 

\begin{figure}
\begin{center}
\includegraphics[width=9cm]{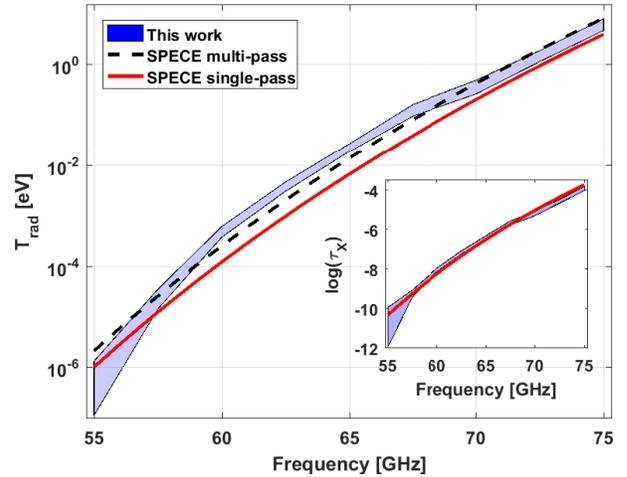}
\caption{Comparison of predicted X-mode ECE spectra from our formalism and from the SPECE code, assuming viewing geometry {\em A}, specular reflection, $R_w=0.65$, and plasma parameters from our ITER baseline scenario. Inset compares our associated ensemble-averaged optical depths to the single-pass values from SPECE.}
\label{fig,spece}
\end{center}
\end{figure}

\section{Discussion}\label{sec,discuss}

Our simulated ECE spectra show a strong dependence on frequency and assumed plasma temperature, along with a modest variation with wall reflectivity. In contrast, model assumptions such as the adopted antenna orientation and angular spread of reflections play a relatively negligible role. Here we will first discuss the assumptions and limitations of our model, before comparing our predictions to other results and considering possible implications.

\subsection{Model assumptions and limitations}\label{sec,assume}
An important unknown in our model is the effective reflectivity $R_w$ of the ITER PFCs, for which we have considered a range of values. In the regime of wall absorption clearly dominating over plasma absorption, Figure~\ref{fig,Trad1} implies a variation in predicted $T_{rad}$ by an average factor of $\sim 2.5$ across
our range of $R_w$. This  uncertainty could potentially be constrained by microwave measurements using mock ITER PFCs combined with raytracing based on a complete 3D vessel model. 
Such a study might also clarify to what extent the complex 3D geometry and surface structure of the ITER PFCs can be handled statistically by assuming randomized reflection around the specular direction as done here. We note here that across the $R_w = 0.8$--0.9 range that likely applies to the metallic ITER vessel at these frequencies, the difference in our model predictions is comparable to the associated statistical uncertainties. This suggests that systematic and statistical uncertainties related to
our treatment of the vessel walls are small and large enough, respectively, to render our conclusions reasonably robust against modest deviations from our assumptions.

We also do not incorporate the dependence of the wall reflectivity on incident angle.
Furthermore, there is no accounting for effects relating to cross-polarized O-mode rays travelling to a different part of the machine and there getting converted back to X-mode. Only a single ray is traced and the reconverted X-mode rays are incorporated back into that ray. Tracing a single ray also implies that we do not account for diffraction effects that may broaden the beam and lower its central intensity, nor for the antenna pattern which determines how the measured ECE signal represents a weighted sum of $T_{rad}$ over all incoming rays. 
However,  averaging over many randomized reflections, and considering that an antenna detects radiation from virtually anywhere in the plasma in our approach,
 these effects are arguably all included in a statistical sense.

A final assumption worth highlighting is the fact that a Maxwellian velocity distribution is assumed for the electrons, but auxiliary heating and energy transfer from fusion alphas may generate a significant non-thermal high-energy tail in the electron distribution function $f_e$ in ITER. Due to relativistic downshifting, such suprathermal electrons could boost the ECE signal measured at the outboard side around optically thin harmonics, including the X-mode ECE in ITER below the fundamental resonance  \cite{born96,born83}. Knowledge of $f_e$ under various operating conditions would be needed to account for this, but, under the adopted assumptions, our results may be viewed as an estimate of the ECE signal in the limit of approximately Maxwellian electron populations.

To explore this assumption further, we estimated the characteristic electron velocities $\beta = v_e/c$ contributing to the ECE signal within our 55--75~GHz frequency range. First, we computed ECE weight functions $w$ \cite{heid07,sale14} using equations~(5.2.16) and (5.2.68) in \cite{hutc02} and averaged these over all 
viewing angles relative to the magnetic field (cf.\ Section~\ref{sec,depend}) and over the field strengths of $B \approx 4.8$--5.8~T at which most of the ECE signal originates according to our raytracing. The origin of the generated ECE signal in electron velocity space was then inferred by multiplying the resulting $w$ with a $T_e = 20$~keV non-relativistic Maxwellian $f_e$. The results suggest that the strongest contribution arises from electrons with $\beta \approx 0.7$, corresponding to 2.5 times the characteristic thermal velocity $v_{e,th}$ for a $T = 20$~keV plasma, while the smallest $\beta$ that can contribute is $\beta \approx 0.5$, equivalent to $\approx1.8v_{e,th}$. Hence, the emission in our model arises from electrons that are still well within the Maxwellian tail rather than being strongly supra-thermal or highly relativistic. The relativistic mass increase and the radiation reaction force (see, e.g., \cite{denk18}) would furthermore reduce the ECE signal from high--$\beta$ electrons compared to our assumed case. We thus find our assumptions of a thermalized electron distribution and a weakly relativistic absorption coefficient reasonably well justified. However, we will aim to include the effect of non-thermal electrons in future iterations of our model and to test the resulting predictions experimentally.

\subsection{Diagnostic implications}

At frequencies below $f=70$~GHz, our results suggest that the X-mode $T_{rad}$ may range from negligible to a significant fraction of the core $T_e$.  This highly frequency-- and temperature dependent behaviour will not be captured by the planned ITER ECE diagnostic but could potentially comprise a useful diagnostic of the plasma temperature in its own right, as suggested by Figure~\ref{fig,Tscaling}. In its simplest form, this would yield a non-localized measurement whose interpretation would depend on assumptions on the density and temperature {\em profiles}.
The sensitivity of ECE at $\omega < \omega_c$ to most of the plasma volume renders such spectra difficult to invert both in real and velocity space, although physically motivated prior information on the electron distribution function, along the lines of \cite{sale16}, might reduce the high dimensionality of the problem. If further input from other diagnostics such as Thomson scattering and X-ray/gamma-ray spectrometry \cite{noce17} could make an inversion tractable using velocity-space tomography, the results could provide spatial information on the electron velocity distribution (or, perhaps more realistically, simply aid in pinpointing deviations from an {\em assumed} distribution function \cite{sale16}) -- independently of measurements in the optically thick regime. This approach could potentially be tested using radiometers operating at $\omega < \omega_c$ in existing fusion devices or using the 55--65~GHz CTS receivers on ITER itself.

For ITER CTS, the primary purpose is to diagnose fast ions, based on detecting scattered 60~GHz probe gyrotron radiation in X-mode. As the CTS noise level scales with $T_{rad}$  (e.g.\ \cite{bind92b}), Figure~\ref{fig,Trad1} shows that the upshifted ($f > 60$~GHz) part of ITER CTS spectra will be affected in particular. This is especially relevant for hotter plasmas, where the predicted keV-level intensities may affect the accuracy with which $\alpha$-particle densities can be recovered with CTS. On the other hand, even across the fairly narrow 55--65~GHz frequency range covered by the CTS receivers and for low absolute ECE levels, the spectrum of the inferred diagnostic background might provide useful temperature estimates.
This could be employed both for calibration purposes and as diagnostic cross-validation of optically thick ECE measurements.

For X-mode reflectometry, the predicted increase in ECE levels with $f$ and $T_e$ may result in relatively better localization of the reflecting layer, and hence improved spatial resolution, for plasma locations corresponding to the low end of our frequency range. Again, this would be particularly pronounced for the higher plasma temperatures associated with the adopted ITER hybrid scenario.

Finally, we note that our results could also impact on the feasibility of alternative uses of planned microwave diagnostics. For example, using the ITER ECE or CTS receivers it might be possible to search for evidence of the theoretically predicted 50--100~GHz hyperfine structure transitions from certain isotopes of plasma impurity ions (e.g.\ \cite{zhan00,godd03,suny07}). 
This would comprise an alternative means of detecting impurities from gas seeding or wall recycling and of testing fundamental nuclear physics. Our results imply that such studies 
should focus on lower-frequency lines and cooler ITER plasmas, where the anticipated ECE background will be far lower.

\section{Summary and outlook}\label{sec,conclude}

The prediction of electron cyclotron emission from fusion devices at frequencies for which the plasma is optically thin is sensitive to assumptions regarding wall reflections.
Instead of carefully accounting for the complex 3D geometry and reflectance properties of the plasma-facing components in a typical toroidal device, we have developed an approach to
simulating ECE spectra in this regime that relies on ensemble-averaging of the results of many randomized reflections. In addition to plasma absorption and emission, the method accounts, in an approximate and statistical manner, for wall reflections, wall absorption, cross-polarization, and deviations from toroidal symmetry in a realistic 
 poloidal geometry.

Our framework can potentially be applied to any toroidal fusion device, and we have here used it
to simulate X-mode ECE spectra for the baseline and hybrid plasma scenarios in ITER at frequencies well below the fundamental EC resonance at the
nominal ITER toroidal field. Given a plasma equilibrium and assuming diffuse reflections, the only free model parameter of any consequence for our results is the assumed wall reflection coefficient.

Our results indicate a strong dependence of the ECE radiation temperature on frequency in the considered interval of 55--75~GHz, increasing by five orders of magnitude for the ITER baseline plasma scenario. In the 55--65~GHz range to be covered by the ITER CTS diagnostic, simulated ECE levels remain  below 10~eV for relevant wall reflectivities, increasing to hundreds of eV around the $\sim 75$~GHz frequencies covered by reflectometry. In the higher-temperature ITER 12.5~MA hybrid scenario, corresponding levels are in the keV range, with potential impact on the performance of both diagnostics and on the accuracy of total radiation losses determined by the planned ECE diagnostic which does not cover frequencies below 70~GHz.
For the benefit of, e.g., diagnostic modellers, we have provided approximate analytical relations, equations~(\ref{eq,analyt1})--(\ref{eq,analyt3}), on the basis of our numerical results for
the baseline scenario. These relations allow straightforward estimation of ITER ECE levels within our considered frequency range as a function of frequency, plasma temperature, and wall reflectivity, as well as of the maximum frequencies at which $T_{rad}$ remains below specific levels.

For our default assumptions using randomized reflections, simulated ECE levels at 60~GHz are two orders of magnitude higher than if assuming specular reflection. This is a consequence of randomized reflections producing ray segments which, on average, have longer and more oblique paths through the plasma core and so can experience significant radiation temperature increments before losing energy in wall reflections. Hence, diffuse reflections associated with deviations from strict toroidal symmetry may significantly elevate the ECE levels in ITER in the optically thin regime.

Self-consistently computing the X-mode conversion coefficient $p_x$, i.e.\ the average fraction of X-mode power transferred to O-mode in wall reflections, we find typical values of $p_x = 0.2$--0.3, in good agreement with empirical estimates from existing fusion devices. Comparison of predicted radiation temperatures in the limit of specular reflection to corresponding results from the SPECE code  -- itself used successfully to model ECE spectra at JET in the optically thick regime -- also reveals good quantitative agreement across the full frequency range, lending confidence in our approach.
 
A systematic uncertainty  in our simulation results  is the assumed average wall reflectivity $R_w$, with predicted ECE levels varying by a factor of $\sim 2.5$ across
the considered range of $R_w=0.6$--0.9. At typical metal wall reflectivities of $R_w \gtrsim 0.8$, this variation is, however, broadly comparable to the statistical uncertainties resulting from our approach. Future empirical and numerical studies should be able to constrain  $R_w$ for the ITER vessel and hence
the range in radiation temperature allowed by our results. Another possible model uncertainty is our assumption of a thermal velocity distribution for the electrons, but consideration of
the characteristic electron velocities contributing to the predicted ECE levels suggests that this is a reasonable approximation. 

Our framework forms a useful basis for 
assessing cyclotron emission from thermalized electrons in ITER in the optically thin regime and the resulting impact on anticipated measurements with microwave diagnostics such as reflectometry, collective Thomson scattering, and the ECE diagnostic itself.
However, our model is not restricted to consideration of strongly downshifted ECE emission in ITER. In principle, it should thus be possible to test the model on existing tokamaks wherever sufficiently sensitive measurements at low optical depth are available. This could validate the physically motivated assumptions underlying our wall reflection model, test additional model assumptions associated with our raytracing (see Section~\ref{sec,assume}), and potentially even allow assessment of the average $R_w$ of various fusion devices at the relevant frequencies. Such a study will be the subject of future work.

\section*{Acknowledgments}
We thank R.~V.~Budny and S.~M.~Kaye for providing discharge parameters for TFTR discharge \#55851.
The work leading to this publication has been funded partially by Fusion for Energy under Grant F4E-FPA-393. This publication reflects the views only of the authors, and Fusion for Energy cannot be held responsible for any use which may be made of the information contained therein.

\end{document}